\documentclass[fleqn,twoside]{article}
\usepackage{espcrc2}

\usepackage{graphicx}
\usepackage[figuresright]{rotating}

\newcommand{\AmS}{{\protect\the\textfont2
  A\kern-.1667em\lower.5ex\hbox{M}\kern-.125emS}}
\newcommand{\sims}{\hspace{.3em}\raisebox{.4ex}{$<$}\hspace{-.75em}
\raisebox{-.7ex}{$\sim$}\hspace{.3em}} 
\newcommand{\siml}{\hspace{.3em}\raisebox{.4ex}{$>$}\hspace{-.75em}
\raisebox{-.7ex}{$\sim$}\hspace{.3em}} 

\title{ CP$^{N-1}$ model with the theta term and maximum entropy method
        \thanks{Talk presented by Y. Shinno}
	\thanks{This work is supported in part by Grants-in-Aid for
        Scientific Research (C)(2) of the JSPS (No. 15540249) and of the
        Ministry of Education, Culture, 
        Sports, Science and Technology( No.'s 13135213 and 13135217).}
	\thanks{SAGA-HE-214, YGHP-04-33}}

\author{ Masahiro Imachi\address[DPYU]{Department of Physics, Yamagata
         University, Yamagata 990-8560, Japan},
	 Yasuhiko Shinno\address[GSSESU]{Graduate School of Science and
         Engineering, Saga University, Saga 840-8502, Japan} and 
	 Hiroshi Yoneyama\address[DPSU]{Department of Physics, Saga
         University, Saga 840-8502, Japan}
}        

\begin{document}

\begin{abstract}
 A $\theta$ term in lattice field theory causes the sign problem in 
 Monte Carlo simulations. This problem
 can be circumvented by Fourier-transforming the topological charge
 distribution $P(Q)$. This strategy, however, has a limitation, because
 errors of $P(Q)$ prevent one from calculating the partition function
 ${\cal Z}(\theta)$ properly for large volumes. This is called
 flattening. As an alternative approach to the Fourier method, we utilize
 the maximum entropy method (MEM) to calculate ${\cal Z}(\theta)$. We
 apply the MEM to Monte Carlo data of the CP$^3$ model. It is found that 
 in the non-flattening case, the result of the MEM agrees with that of the
 Fourier transform, while in the flattening case, the MEM gives smooth
 ${\cal Z}(\theta)$.
\vspace{1pc}
\end{abstract}

\maketitle

\section{INTRODUCTION}
 It is well known that QCD  in principle has a $\theta$ term. The
 $\theta$ term is deeply associated with non-perturbative properties of
 QCD at low energy and provides us with interesting issues such as the
 strong CP problem, possibilities of rich phase structures in $\theta$
 space and so on. So it is a challenging subject to investigate the
 dynamics of QCD with the $\theta$ term. \par
 The $\theta$ term in lattice field theory makes the Boltzmann weight
 complex and prevents one from performing Monte Carlo (MC) simulations 
 directly. This is the sign problem. This problem
 can be circumvented by Fourier-transforming the topological charge
 distribution\cite{rf:BRSW,rf:Wiese,rf:BISY}
\begin{equation}
 P(Q)\equiv\frac{\int[{\cal D}\phi]_Q\;e^{-S(\phi)}}
  {\int{\cal D}\phi\;e^{-S(\phi)}},
\end{equation}
where $Q,\;\phi\;{\mbox {\rm and}}\;S$ are the topological charge, a
field of the system and an action, respectively. The measure $[{\cal
D}\phi]_Q$ implies that the integral is restricted to configurations of
$\phi$ with $Q$. The partition function is given in terms of $P(Q)$ by
\begin{equation}
 {\cal Z}(\theta)=\sum_Q P(Q)e^{i\theta Q}.
\end{equation}
%
\par
 Although this method works well for small volumes, it does not work for
 large ones because errors of $P(Q)$ affect strongly the behavior of the
 free energy density, $f(\theta)\equiv -\frac{1}{V}\log{\cal Z}(\theta)$
 ($V$ is a volume). This is called flattening\cite{rf:PS,rf:IKY}. Flattening
 can be remedied by 
 reducing the errors, but this is hopeless, because exponentially
 increasing statistics are needed as volume increases. \par
 In order to deal with flattening, we have utilized the maximum entropy
 method (MEM)\cite{rf:Bryan,rf:JG,rf:AHN}. In ref.\cite{rf:ISY}, we
 applied the MEM to mock data of the Gaussian $P(Q)$ to study whether
 the MEM is effective to our issue. In the non-flattening case, the MEM
 reproduces the exact ${\cal Z}(\theta)$, while it 
 gives smooth ${\cal Z}(\theta)$ in the case where flattening occurs in
 the Fourier method.
 In the present work, as the next step, we 
 apply the MEM to MC data of the CP$^{N-1}$ model, which has several
 dynamical properties in common with QCD.

\section{FLATTENING AND MEM}
 We simulated the CP$^3$ model with a fixed point action at a fixed
 coupling constant, $\beta=3.0$\cite{rf:BISY}. The lattice extension $L$
 is changed from 4 to 96. The statistics are more than 1 million 
 for each case. \par
\begin{figure}[htb]
\vspace*{-3mm}
\centerline{\includegraphics[width=80mm, height=55mm]{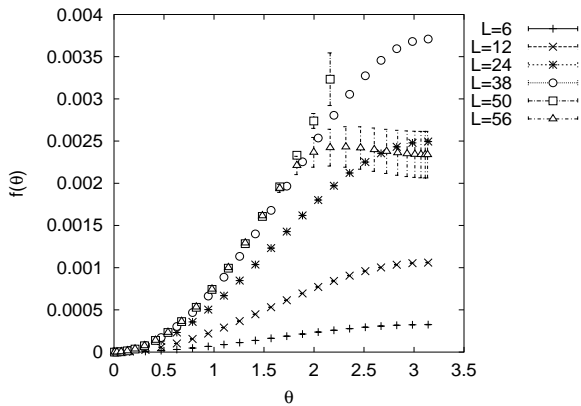}}
\vspace*{-8mm}
\caption{Free energy density $f(\theta)$ of the MC data of the CP$^3$
 model. Lattice extension $L$ of the system is changed from 6 to 56.}
\vspace*{-2mm}
\label{fig:free}
\end{figure}
 Figure~\ref{fig:free} displays $f(\theta)$ obtained numerically by
 Fourier-transforming $P(Q)$ of the MC data for various volumes. Up to
 $L=38$, the Fourier transform works, but $f(\theta)$ for
 $L=50$ and 56 cannot be calculated properly. Especially for $L=56$,
 $f(\theta)$ becomes flat for $\theta\siml 2.0$. This is
 nothing but flattening. The density $f(\theta)$ for $L=50$ breaks down
 for $\theta\siml 2.3$ due to the negative values of ${\cal
 Z}(\theta)$. We also call it flattening because
 errors of $P(Q)$ disturb the behavior of $f(\theta)$.
\vspace*{3mm}\par 
 In order to deal with flattening, we employ the MEM. To this end, we
 utilize the inverse Fourier transform. The MEM is based on Bayes'
 theorem and gives the most probable image of ${\cal Z}(\theta)$.
 In our case, the most
 important object is the posterior probability ${\rm prob}({\cal
 Z}(\theta)|P(Q),I)$ which
 is the probability that ${\cal Z}(\theta)$ is realized when the MC data
 of $\{P(Q)\}$  and information $I$ are given. Information $I$
 represents our state of knowledge about ${\cal Z}(\theta)$. Here we
 impose the criterion ${\cal Z}(\theta)>0$. \par 
 The probability ${\rm prob}({\cal Z}(\theta)|P(Q),I)$ is represented in
 terms of $\chi^2$ and the entropy $S$: 
\begin{equation}
 {\rm prob}({\cal Z}(\theta)|P(Q),I)\propto 
  \exp\biggl[-\frac{1}{2}\chi^2 + \alpha S\biggr],
\end{equation}
 where $\alpha$ is a real positive parameter. Conventionally the
 Shannon-Jaynes entropy is employed;
\begin{equation}
 S=\int_{-\pi}^{\pi}d\theta\biggl[{\cal Z}(\theta)-m(\theta)-
  {\cal Z}(\theta)\log\frac{{\cal Z}(\theta)}{m(\theta)}\biggr],
\end{equation}
 where $m(\theta)$ is called the default model and is chosen so as to be
 consistent with information $I$. To sum up, our task is to calculate
 the most probable image ${\hat {\cal Z}}(\theta)$ such that ${\rm
 prob}({\cal Z}|P(Q),I)$ is maximized( see refs.\cite{rf:AHN,rf:ISY} for
 details).

\section{RESULTS}
 We apply the MEM to such MC data that flattening occurs in the Fourier
 method as well as to such MC data that it does not. Here we use the MC data
 for $L=38$(data A) as an example of the non-flattening case and for
 $L=50$(data B) for the flattening one. Two types of default model
 are used: (i) Gaussian type,
 $m_G(\theta)=\exp\bigl[-\gamma\frac{\log10}{\pi^2}\theta^2\bigr]$,
 where a parameter $\gamma$ is changed from 0.6 to 5.0. (ii) weak
 coupling region type, $m_{\rm w}(\theta)$, which will be explained later. 
 In the analysis, the Newton
 method with quadruple precision is used to calculate an image ${\hat
 {\cal Z}}(\theta)$ with high precision. \par
\begin{figure}[htb]
\vspace*{-3mm}
\centerline{\includegraphics[width=65mm, height=60mm]{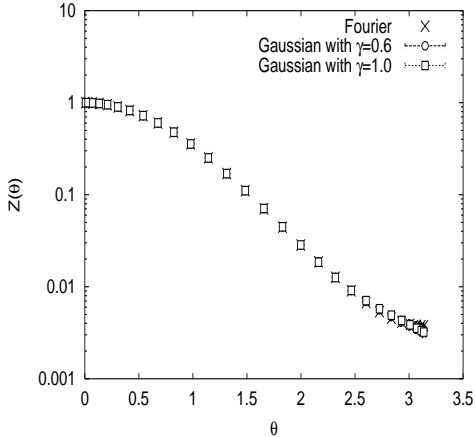}}
\vspace*{-8mm}
\caption{ The most probable image ${\hat {\cal Z}}(\theta)$ of the MC
 data for $L=38$. The Gaussian type default models with $\gamma=0.6$ and
 1.0 are used. As a comparison, the result of the Fourier
 transform($\times$) is also displayed.} 
\vspace*{-2mm}
\label{fig:dataA}
\end{figure}
 Figure~\ref{fig:dataA} displays the most probable image ${\hat {\cal
 Z}}(\theta)$ of data A, as an example of the non-flattening
 case. The Gaussian type default models $m_G(\theta)$ with $\gamma=0.6$ and
 1.0 are used. The partition function ${\cal Z}(\theta)(\equiv{\cal
 Z}_F(\theta))$ obtained by Fourier-transforming $P(Q)$ of data A is
 also displayed.  The result
 of the MEM agrees with that of the Fourier transform 
 in the whole $\theta$ region. Note that errors of ${\hat {\cal Z}}(\theta)$
 are too small to be visible, where the error of ${\hat {\cal
 Z}}(\theta)$ means the uncertainty of ${\hat {\cal
 Z}(\theta)}$\cite{rf:JG,rf:AHN}.  \par
\begin{figure}[htb]
\vspace*{-3mm}
\centerline{\includegraphics[width=65mm, height=60mm]{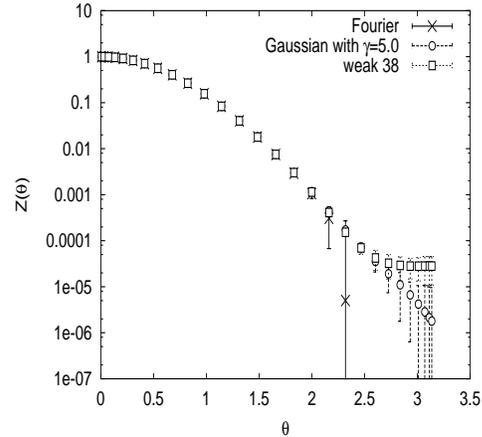}}
\vspace*{-8mm}
\caption{ The most probable image ${\hat {\cal Z}}(\theta)$ of the MC
 data for $L=50$. Two types of default model are used. The result of the
 Fourier transform($\times$) is also displayed.} 
\vspace*{-2mm}
\label{fig:dataB}
\end{figure}
 Next, let us turn to the data B, the flattening case. The results are
 shown in Fig.~\ref{fig:dataB}.  The partition function ${\cal
 Z}_F(\theta)$ of data B breaks down for $\theta\siml 2.3$ because the
 values of ${\cal Z}_F(\theta)$ become negative in this region. For the
 MEM analysis, the Gaussian type $m_G(\theta)$ with
 $\gamma=5.0$ and the weak coupling region type $m_{\rm w}(\theta)$ are
 used. Since $P(Q)$ of data A was successfully Fourier-transformed to
 ${\cal Z}(\theta)\equiv{\cal Z}_A(\theta)$, we use ${\cal Z}_A(\theta)$
 as $m_{\rm w}(\theta)$ for the analysis of data for larger volumes. The
 results of the MEM reproduce smooth 
 ${\hat{\cal Z}}(\theta)$ in the whole $\theta$ region.
 For $\theta\siml 2.7$ they depend on the default models and give no
 definite solution about the behavior of ${\cal Z}(\theta)$ for
 $\theta\simeq\pi$ due to the large magnitude of the errors. 
 For $\theta\sims 2.7$, however, the results of the MEM
 show no $m(\theta)$-dependence and their errors are small.
 We thus obtain a reasonably good solution of ${\cal Z}(\theta)$ for
 $L=50$ up to $\theta=2.7$, which is wider than the valid region of the
 Fourier method.

\section{SUMMARY}
 We applied the MEM to the MC data of the CP$^3$ model for
 various volumes. It was found that the MEM is applicable to the CP$^3$ 
 model in the non-flattening and the flattening cases, and that the
 MEM has the advantage of calculating ${\cal Z}(\theta)$ for somewhat
 wider $\theta$ region than the Fourier transform at least for $L=50$. 
 The following subjects remain to be investigated; (i) a systematic
 study of the $m(\theta)$-dependence of ${\hat {\cal Z}}(\theta)$ and
 its error, (ii) investigation on a question whether the MEM is more
 effective than the Fourier method, and (iii) a study of the feasibility
 of distinction between flattening and the 
 signal of a first order phase transition\cite{rf:ISY2}.

\end{document}